\documentclass[epj]{svjour}
\usepackage{graphics}
\begin{document}
\title{On the nature of the lowest $1/2^-$ baryon nonet and decuplet}
\author{B. S. Zou
}                     
\institute{Institute of High Energy Physics, CAS, Beijing 100049,
China \\ Theoretical Physics Center for Science Facilities, CAS,
Beijing 100049, China}
\date{Received: date / Revised version: date}
%
\abstract{From recent study of properties of the lowest spin-parity
$1/2^-$ baryons, $N^*(1535)$ and $\Delta^*(1620)$, new pictures for
the internal structure of the lowest $1/2^-$ baryon octet and
decuplet are proposed. While the lowest $1/2^-$ baryon octet may
have large diquark-diquark-antiquark component, the lowest $1/2^-$
baryon decuplet is proposed to have large vector-meson-baryon
components. Evidence for the ``missing" members of the new pictures
is pointed out and suggestions are made for detecting these
predicted states from forthcoming experiments.
} 
\maketitle
\section{Introduction}
\label{intro} The classical simple 3q constituent quark model has
been very successful in explaining the static properties, such as
mass and magnetic moment, of the spatial ground states of the flavor
SU(3) octet and decuplet baryons. Its predicted $\Omega$ baryon with
mass around 1670 MeV was discovered by later experiments. However
its predictions for the spatial excited baryons are not so
successful. In the simple 3q constituent quark model, the lowest
spatial excited baryon is expected to be a ($uud$) $N^*$ state with
one quark in orbital angular momentum $L=1$ state, and hence should
have negative parity. Experimentally \cite{PDG06}, the lowest
negative parity $N^*$ resonance is found to be $N^*(1535)$, which is
heavier than two other spatial excited baryons : $\Lambda^*(1405)$
and $N^*(1440)$. In the classical 3q constituent quark model, the
$\Lambda^*(1405)$ with spin-parity $1/2^-$ is supposed to be a
($uds$) baryon with one quark in orbital angular momentum $L=1$
state and about 130 MeV heavier than its $N^*$ partner $N^*(1535)$;
the $N^*(1440)$ with spin-parity $1/2^+$ is supposed to be a ($uud$)
state with one quark in radial $n=1$ excited state and should be
heavier than the $L=1$ excited ($uud$) state $N^*(1535)$, noting the
fact that for a simple harmonic oscillator potential the state
energy is $(2n+L+3/2)\hbar\omega$. So for these three lowest spatial
excited baryons, the classical quark model picture is already
failed.

Evidence is accumulating for the existence of significant intrinsic
non-perturbative 5-quark components in baryons \cite{zou07}. A
well-established fact from electron-proton deep inelastic scattering
and Drell-Yan process is that in the proton the number of $\bar d$
is more than $\bar u$ by an amount $\bar d-\bar u\approx 0.12$
\cite{Garvey}. This obviously cannot be explained by the classical
quark models, but can be easily explained by a mixture of
$n(udd)\pi^+(u\bar d)$ in meson-cloud model \cite{Thomas} or
$[ud][ud]\bar d$ penta-quark configuration \cite{zr}. If there are
already significant 5-quark components in the proton, we would
expect more significant 5-quark components in excited baryons.

To understand the full baryon spectroscopy, it is crucial to
understand the lowest $1/2^-$ baryon nonet and decuplet first !

\section{Nature of $N^*(1535)$ and its $1/2^-$ nonet partners}
\label{sec:1}

\begin{figure}
\resizebox{0.5\textwidth}{!}{%
  \includegraphics{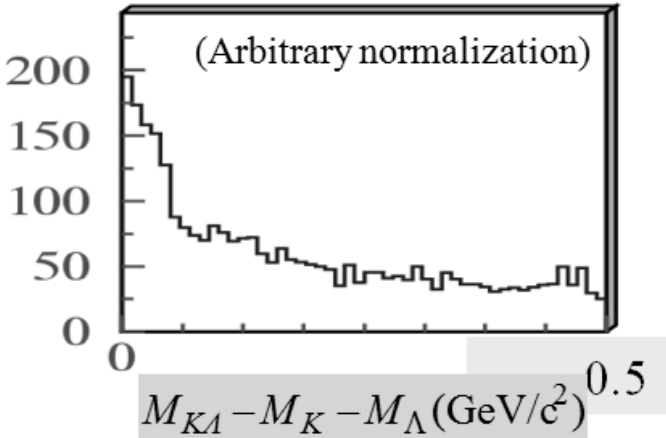}}
\caption{Invariant mass spectrum divided by efficiency and phase
space vs $M_{K\Lambda}\!-\!M_K\!-\!M_\Lambda$ for $J/\psi\to
pK^-\bar\Lambda+c.c.$ \cite{Yanghx}.}
\label{fig:1}       
\end{figure}

Recently BES experiment at Beijing Electron-Positron Collider (BEPC)
has been producing very useful information on $N^*$ resonances
\cite{ppeta,Yanghx,pnpi,weidh}. In $J/\psi\to\bar pp\eta$, as
expected, the $N^*(1535)$ gives the largest contribution
\cite{ppeta}. In $J/\psi\to pK^-\bar\Lambda+c.c.$, a strong
near-threshold enhancement is observed for $K\Lambda$ invariant mass
spectrum \cite{Yanghx} as duplicated in Fig.~\ref{fig:1}. The
$K\Lambda$ threshold is 1609 MeV. The near-threshold enhancement is
confirmed by $J/\psi\to nK_S\bar\Lambda+c.c.$ \cite{weidh}. Since
the mass spectrum divided by efficiency and phase space peaks at
threshold, it is natural to assume it comes from the sub-threshold
nearby $N^*(1535)$ resonance. Then from BES branching ratio results
on $J/\psi\to\bar pp\eta$ \cite{ppeta} and $\psi\to
pK^-\bar\Lambda+c.c.$ \cite{Yanghx}, the ratio between effective
coupling constants of $N^*(1535)$ to $K\Lambda$ and $p\eta$ is
deduced to be  \cite{lbc}
$$g_{N^*(1535)K\Lambda}/g_{N^*(1535)p\eta}
=1.3\pm 0.3 .$$ With previous known value of $g_{N^*(1535)N\eta}$,
the obtained new value of $g_{N^*(1535)K\Lambda}$ is shown to
reproduce recent $pp\to pK^+\Lambda$ near-threshold cross section
data \cite{cosy} as well. There are also indications for the large
$g_{N^*(1535)K\Lambda}$ from partial wave analysis of $\gamma p\to
K\Lambda$ reactions \cite{Lee}. Taking into account this large
$N^*K\Lambda$ coupling in the coupled channel Breit-Wigner formula
for the $N^*(1535)$, its Breit-Wigner mass is found to be around
1400 MeV, much smaller than previous value of about 1535 MeV
obtained without including its coupling to $K\Lambda$. There is also
evidence for large $g_{N^*(1535)N\eta^\prime}$ coupling from $\gamma
p \to p\eta^\prime$ reaction at CLAS \cite{etap}, and large
$g_{N^*(1535)N\phi}$ coupling from $\pi^- p \to n\phi$ and $pp\to
pp\phi$ reactions \cite{xiejj}, but smaller coupling of
$g_{N^*(1535)K\Sigma}$ from comparison of $pp\to p K^+\Lambda$ to
$pp \to p K^+\Sigma^0$ \cite{Sibir}.

The nearly degenerate mass for the $N^*(1535)$ and the $N^*(1440)$
resonances can be easily understood by considering 5-quark
components in them \cite{lbc,zhusl}. The $N^*(1535)$ could be the
lowest $L=1$ orbital excited $|uud>$ state with a large admixture of
$|[ud][us]\bar s>$ pentaquark component having $[ud]$, $[us]$ and
$\bar s$ in the ground state.  The $N^*(1440)$ could be the lowest
radial excited $|uud>$ state with a large admixture of
$|[ud][ud]\bar d>$ pentaquark component having two $[ud]$ diquarks
in the relative P-wave. While the lowest $L=1$ orbital excited
$|uud>$ state should have a mass lower than the lowest radial
excited $|uud>$ state, the $|[ud][us]\bar s>$ pentaquark component
has a higher mass than $|[ud][ud]\bar d>$ pentaquark component. The
large mixture of the $|[ud][us]\bar s>$ pentaquark component in the
$N^*(1535)$ may also explain naturally its large couplings to the
$N\eta$, $N\eta^\prime$ and $K\Lambda$ meanwhile small couplings to
the $N\pi$ and $K\Sigma$. In the decay of the $|[ud][us]\bar s>$
pentaquark component, the $[ud]$ diquark with isospin $I=0$ is
stable and keeps unchanged while the $[us]$ diquark is broken to
combine with the $\bar s$ to form either $K^+(u\bar
s)\Lambda([ud]s)$ or $\eta(s\bar s)p([ud]u)$.

The lighter $\Lambda^*(1405)1/2^-$ is also understandable in this
picture. Its main 5-quark configuration is $|[ud][us]\bar u>$ which
is lighter than the corresponding 5-quark configuration
$|[ud][us]\bar s>$ in the $N^*(1535)1/2^-$.

If this picture of large 5-quark mixture is correct, there should
also exist the SU(3) nonet partners of the $N^*(1535)$ and
$\Lambda^*(1405)$, {\sl i.e.}, an additional $\Lambda^*~1/2^-$
around 1570 MeV, a triplet $\Sigma^*~1/2^-$ around 1360 MeV and a
doublet $\Xi^*~1/2^-$ around 1520 MeV \cite{zhusl}. There is no hint
for these baryon resonances in the PDG tables \cite{PDG06}. However,
as pointed out in Ref.\cite{zou07}, there is in fact evidence for
all of them in the data of $J/\psi$ decays. According to PDG
\cite{PDG06}, the branching ratios for
$J/\psi\to\bar\Sigma^-\Sigma^*(1385)^+$ and
$J/\psi\to\bar\Xi^+\Xi^*(1530)^-$ are $(3.1\pm 0.5)\times 10^{-4}$
and $(5.9\pm 1.5)\times 10^{-4}$, respectively. These two processes
are SU(3) breaking decays since $\Sigma$ and $\Xi$ belong to SU(3)
$1/2^+$ octet while $\Sigma^*(1385)$ and $\Xi^*(1530)$ belong to
SU(3) $3/2^+$ decuplet. Comparing with the similar SU(3) breaking
decay $J/\psi\to\bar p\Delta^+$ with branching ratio of less than
$1\times 10^{-4}$ and the SU(3) conserved decay $J/\psi\to\bar
pN^*(1535)^+$ with branching ratio of $(10\pm 3)\times 10^{-4}$, the
branching ratios for $J/\psi\to\bar\Sigma^-\Sigma^*(1385)^+$ and
$J/\psi\to\bar\Xi^+\Xi^*(1530)^-$ are puzzling too high. A possible
explanation for this puzzling phenomena is that there were
substantial components of $1/2^-$ under the $3/2^+$ peaks but the
two branching ratios were obtained by assuming pure $3/2^+$
contribution. This possibility should be easily checked with the
high statistics BESIII data in near future.

\section{Nature of $\Delta^{++*}(1620)$ and its $1/2^-$ decuplet partners}

The spectrum of isospin 3/2 $\Delta^{++*}$ resonances is of special
interest since it is the most experimentally accessible system
composed of 3 identical valence quarks. However, our knowledge on
these resonances mainly comes from old $\pi N$ experiments and is
still very poor~\cite{PDG06}. A possible new excellent source for
studying $\Delta^{++*}$ resonances is $pp \to nK^+\Sigma^+$
reaction, which has a special advantage for absence of complication
caused by $N^*$ contribution because of the isospin and charge
conversation.

\begin{figure}
\resizebox{0.5\textwidth}{!}{%
  \includegraphics{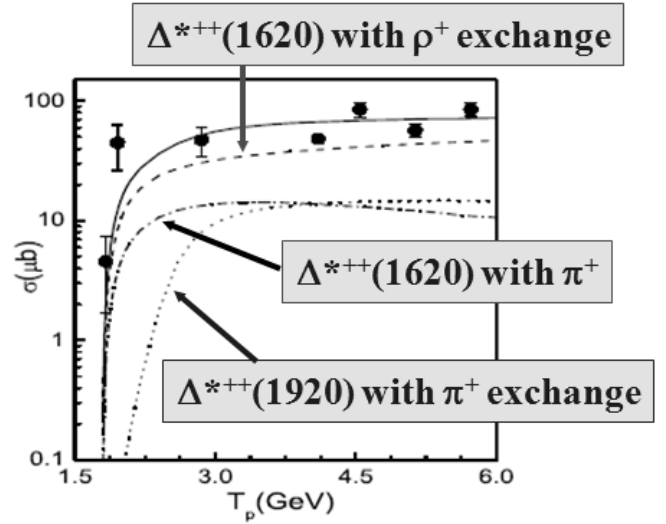}}
\caption{Total cross section vs kinetic energy of proton beam for
the $pp \to nK^+\Sigma^+$ reaction: data~\cite{data,06cosy11} and
calculation (solid curve for sum of other curves) \cite{xiejj2}.}
\label{fig:2}       
\end{figure}

At present, little is known about the $pp\to nK^+\Sigma^+$ reaction.
Experimentally there are only a few data points about its total
cross section versus energy~\cite{data,06cosy11}. Theoretically a
resonance model with an effective intermediate $\Delta^{++*}(1920)$
resonance~\cite{tsushima} and the J\"{u}lich  meson exchange
model~\cite{gas} reproduce the old data at higher beam
energy~\cite{data} quite well, but their predictions for the cross
sections close to threshold fail by order of magnitude compared with
very recent COSY-11 measurement~\cite{06cosy11}. Recently this
reaction was restudied \cite{xiejj2}. With an effective Lagrangian
approach, contributions from a previous ignored
sub-$K^+\Sigma^+$-threshold resonance $\Delta^{++*}(1620)1/2^-$ are
fully included in addition to those already considered in previous
calculations. It is found that the $\Delta^{++*}(1620)$ resonance
gives an overwhelmingly dominant contribution for energies very
close to threshold, with a very important contribution from the
t-channel $\rho$ exchange as shown in Fig.~\ref{fig:2}. This solves
the problem that all previous calculations seriously underestimate
the near-threshold cross section by order(s) of magnitude.

Meanwhile the extra-ordinary large coupling of the
$\Delta^{*}(1620)$ to $\rho N$ obtained from the $\pi^+ p\to
N\pi\pi$ \cite{PDG06,dytman} seems confirmed by the new study
\cite{xiejj2} of the strong near-threshold enhancement of $pp \to
nK^+\Sigma^+$ cross section. Does the $\Delta^{*}(1620)$ contain a
large $\rho N$ molecular component or relate to some $\rho N$
dynamical generated state? If so, where to search for its SU(3)
decuplet partners? Sarkar et al.~\cite{oset} have studied baryonic
resonances from baryon decuplet and psudoscalar meson octet
interaction. It would be of interests to study baryonic resonances
from baryon octet and vector meson octet interaction. In fact, from
PDG compilation \cite{PDG06} of baryon resonances, there are already
some indications for a vector-meson-baryon SU(3) decuplet. While the
$\Delta^{*}(1620)1/2^-$ is about 85 MeV below the $N\rho$ threshold,
there is a $\Sigma^*(1750)1/2^-$ about 70 MeV below the $NK^*$
threshold and there is a $\Xi^*(1950)?^?$ about 60 MeV below the
$\Lambda K^*$ threshold. If these resonances are indeed the members
of the $1/2^-$ SU(3) decuplet vector-meson-baryon S-wave states, we
would expect also a $\Omega^* 1/2^-$ resonance around 2160 MeV. All
these baryon resonances can be searched for in high statistic data
on relevant channels from vector charmonium decays by upcoming BES3
experiments in near future.

\section{Conclusion}
While the classical 3q constituent quark model works well in
reproducing properties of baryons in the spatial ground states, the
study of $1/2^-$ baryons seems telling us that the $\bar qqqqq$ in
S-state is more favorable than $qqq$ with $L=1$. In other words, for
excited baryons, the excitation energy for a spatial excitation
could be larger than to drag out a $q\bar q$ pair from gluon field.

Whether the $\bar qqqqq$ components are in penta-quark configuration
or meson-baryon configuration depends on the strength of relevant
diquark or meson-baryon correlations.

For $N^*(1535)$ and its $1/2^-$ SU(3) nonet partners, the diquark
cluster picture for the penta-quark configuration gives a natural
explanation for the longstanding mass-reverse problem of
$N^*(1535)$, $N^*(1440)$ and $\Lambda^*(1405)$ resonances as well as
the unusual decay pattern of the $N^*(1535)$ resonance. Its
predictions of the existence of an additional $\Lambda^*~1/2^-$
around 1570 MeV, a triplet $\Sigma^*~1/2^-$ around 1360 MeV and a
doublet $\Xi^*~1/2^-$ around 1520 MeV \cite{zhusl} could be examined
by forth coming experiments at BEPC2, CEBAF, JPARC etc..

For $\Delta^{*++}(1620)$ and its $1/2^-$ SU(3) decuplet partners,
their SU(3) quantum numbers do not allow them to be formed from two
good scalar diquarks plus a $\bar q$. Then their $\bar qqqqq$
components would be mainly in the meson-baryon configuration. This
picture can be also examined by forth coming experiments.

\bigskip
\noindent
{\bf Acknowledgements} I would like to B.C.Liu, J.J.Xie
and H.C.Chiang for collaborations on relevant issues. This work is
partly supported by the National Natural Science Foundation of China
under grants Nos. 10435080, 10521003 and by the Chinese Academy of
Sciences under project No. KJCX3-SYW-N2.

\end{document}